**Decomposition of Methane Diluted with Inert Gas in an RF Discharge Cell**

S. Gershman[1*], M.N. Shneider[2], Y. Raitses[1]

[1]Princeton Plasma Physics Laboratory, Princeton, NJ

[2]Department of Aerospace and Mechanical Engineering, Princeton University, Princeton, NJ

*sgershma@pppl.gov

**Abstract:** Decomposition of methane using non-thermal plasmas is an attractive route for producing hydrogen-rich gases and valuable carbon nanomaterials. Understanding how plasma discharge modes influence methane decomposition in optimizing plasma-assisted chemical conversion remains unexplored. This study explores the coupling between the discharge structure and product selectivity in RF capacitively coupled discharges operating in methane/inert gas mixtures in the pressure range of 2 – 3 torr. The discharge exhibits mode transitions from uniform to striated in Ar and Kr and from diffuse to contracted in Ar and Kr with <5% $CH_4$. The discharges in He and Ne remained uniform under our operating conditions, and their mixtures with $CH_4$ remained diffuse. A 0-d model for Ar/$CH_4$ discharge established a threshold for contraction while also asserting the importance of $Ar_m^*$ in the dissociation and ionization processes. The highest degree of methane decomposition, >99.7% with the main products of acetylene and graphitized solid carbon was achieved in the contracted discharge mode for both Kr or Ar with ≤ 5% $CH_4$. We demonstrate that contraction can play a crucial role in the effective decomposition of methane with value-added products and that both the electronic and thermal properties of plasma gas are responsible for this effect.

Ethical Compliance: This study did not use human or animal subject or any animal derived products.

Data Access Statement: All relevant data is included in the paper and supplementary information. Any additional data is available upon request.

Conflict of Interest declaration: The authors declare that they have no affiliations with or involvement in any organization or entity with any financial interest in the subject matter or materials discussed in this manuscript.

**Introduction**

Plasma decomposition of methane is a promising pathway for producing valuable solid and gaseous products with applications in the chemical and energy industries, including hydrogen production as a $CO_2$-free alternative to conventional processes.[1,2,3] It is also important in nuclear fusion, where tritiated methane ($CH_3T$) forms in plasma–wall interactions and must be efficiently decomposed to mitigate tritium release.[4,5] Methane containing plasmas enable continuous catalyst-free tunable synthesis of carbon nanomaterials, including nanodiamods and carbon nano-onions, which have been reported in RF, arc, and microwave plasmas.[6–13] Raman spectroscopy and transmission electron microscopy (TEM) indicate the possibility of diamond nucleation in the gas phase within plasma environments.[8,12,13] Other crystalline nanostructures, such as graphitic onion-like carbon nanostructures, have been produced in semi-continuous flow reactors using thermal plasma.[10,11]

The mechanisms of methane decomposition and the role of the plasma in this process remain debatable. In high power arc and microwave plasmas, where gas temperatures exceed 2000 K, $CH_4$ decomposition is largely thermal, driven by pyrolysis at temperatures above 1200 K.[1,2] In contrast, in non-thermal RF plasmas, electrons gain high energies from the oscillating electric field and transfer energy selectively via excitation, dissociation, and ionization processes.[14–18] In such conditions, vibrational excitation of $CH_4$ lowers the effective dissociation barrier. Vibrational excitation can occur either by electron impact or through energy transfer from metastable atoms.[19–21] This highlights the role of inert gas metastables (e.g., $Ar_m$, $He_m$) not only in enhancing plasma density through stepwise and Penning ionization, but also in directly promoting methane dissociation.[18] Thus, understanding how the choice of background inert gas modifies the concentration of metastables and the subsequent pathways of $CH_4$ decomposition is essential for controlling both hydrogen production and carbon nanostructure synthesis.

In this study, we investigated the decomposition of $CH_4$ in RF capacitively coupled plasma at moderate pressures of several torr in methane or methane diluted with inert gases (e.g., He, Ne, Ar, or Kr). We correlated the gas and solid product selectivity with the plasma mode and the choice of the background gas, and addressed the possible reasons for the observed changes in the discharge mode.

The remainder of this paper is organized as follows. Section II covers the experimental setup (Sec. II.1), and experimental methods (Sec. II.2 – II.3) for electrical measurements and optical emission spectroscopy (OES). Section II.3 describes the branching method based on the line emission spectrum that is used to determine the concentrations of the Ar 1s5 and 1s3 metastable states. Experimental results with minimal comments are reported in Section III and organized according to the measurement method, electrical (Sec. III.1), OES (Section III.2), and FTIR gas analyses (Se. III.3), and fast-frame imaging (Sec. III.4), and solid sample analysis (Sec. III.5). The discussion in Section IV is dedicated to the analysis of the contracted discharge mode by 0d modeling (Section IV.1) and an explanation of the temperature differences between the background gases using a simplified energy balance (Section IV.2), which allows us to connect the discharge modes and background gases to the differences in methane decomposition, gaseous, and solid products (Sec. IV.3). We conclude the paper with a few final points (Sec. V).

**II. Experimental Methods**

**II.1. Experimental Setup.**
The experimental setup consisted of a quartz tube with an inner diameter of 10 mm positioned horizontally. The copper ring electrodes were wrapped around the tube, 35 mm apart (Figure 1). The total gas flow rate was maintained at 50 sccm for all the gas mixtures and pure gases using Alicat® mass flow controllers. A mechanical vacuum pump was used to maintain the gas pressure in the range 2.4 – 2.8 torr. Experiments were conducted for pure $CH_4$ and two-gas mixtures of $CH_4$ with $H_2$, He, Ne, Ar, or Kr, with the bulk of the study conducted with Ar/$CH_4$ mixtures. Each gas alone was tested at similar input powers, pressures, and flow rates.

## II.2. Electrical Measurements.

A commercial Kurt Lesker RF Model R301 MKII, power source set to 13.56 MHz and continuous operation was used to power the plasma. A bidirectional (dual directional) coupler was inserted between the power supply and the impedance matching box. The bidirectional coupler measured the forward and reflected signals, which were reduced by 40 dB. The reflected signal was monitored to improve impedance matching. Using the forward and reflected signals from the directional coupler, the reflected power was estimated at <10% for all experiments. The current to the ground from the ground electrode was measured using a Pearson® 2877 coil, and the voltage at the high-voltage electrode was measured using a Tektronix® 6015A passive high voltage probe. Voltage measurements provide information regarding discharge initiation and maintenance. Current measurements were used to estimate plasma density based on the current density. The phase shift between the voltage and current provides information about the RF discharge mode and its capacitive or resistive nature.

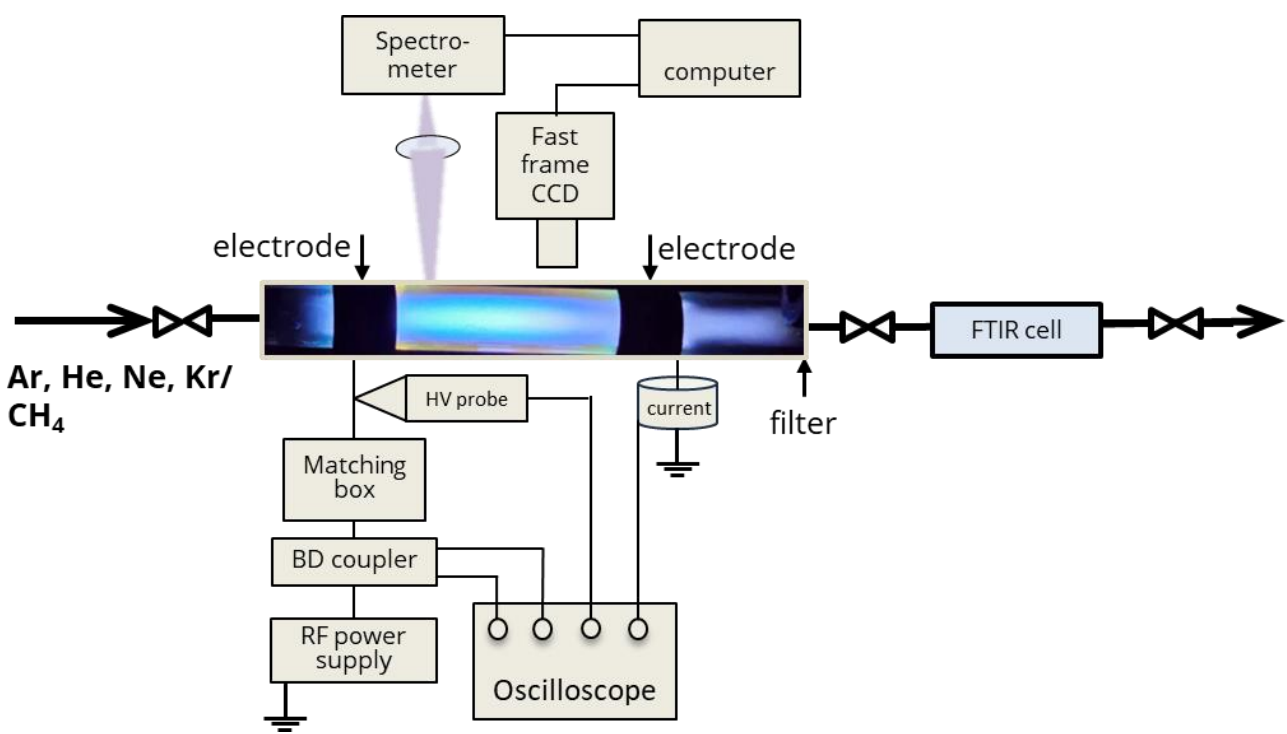

**Figure 1.** Diagram of the experimental setup and diagnostics including fast frame imaging optical emission spectroscopy, FTIR downstream sampling, and electrical measurements.

Outflow gas products were quantified by FTIR absorption (Figure 1) spectroscopy using a 10 cm gas cell with KBr windows and a JASCO FTIR 6800 spectrometer with a nitrogen-cooled mercury-cadmium-tellurite (MCT) detector. The solid products were collected from the tube walls or from the in-flow filter for ex situ analysis using TEM and SEM. The solid samples were analyzed at the Princeton Material Science Institute.

Optical emission was recorded using a broadband Ocean Optics® HR6 spectrometer (190 – 1100 nm range, 0.71 nm resolution). A fast frame camera, Phantom 4 (Vision Research), was used for fast imaging of the plasma instabilities. Because the frame rate and exposure time were kept constant for easier comparison, some images were oversaturated owing to the 8-bit camera depth. The maximum frame rate used was 10,000 fps with a 10 µs exposure time.

## II.3. Optical Emission Spectroscopy Measurement and Analysis

A broadband emission spectrum was recorded using an HR6 OceanOptics® spectrometer with an integration time of 10 ms. The spectrum was used to monitor the emission from the excited species of interest, $C_2$, $H_2$, CH, and the atomic emission for H, Ar, He, Ne, and Kr. Intensity calibration was performed using UV-VIS radiometric calibration light sources in the 280 – 1200 nm range. This allows the analysis of the line intensity ratios for lines in different parts of the spectrum, such as the hydrogen Balmer series.

In low-temperature RF plasmas with $CH_4$ added to inert gases, electron-impact excitation (including dissociative excitation) populates electronic, vibrational, and rotational states such as CH(A–X), and the de-excitation of these states produces the $C_2$ Swan system, routinely analyzed by OES.[22–25] At pressures around ~2.5 Torr, rotational excitations can equilibrate with the gas on nanosecond time scales, so rotational temperatures inferred from OES can often serve as proxies for the gas temperature.[23] Vibrational–translational relaxation is much slower and hence can maintain non-equilibrium with the gas temperature.[22–24] Typically, in non-equilibrium RF plasmas Te >> Tvib > Trot, electrons have much higher mean energy that drives excitation and dissociation. In this work we use OES-Toolbox[26] (Julian Held (2024) OES-toolbox: v0.3.2 https://doi.org/10.5281/zenodo.13986865 ) to estimate the $T_{rot}$ and $T_{vib}$ using $C_2$ emission.[27] The software uses a collisional radiative model to predict the rotational and vibrational excitation temperature for $C_2$ emission.

When available in the spectrum, $H_\alpha$ and $H_\beta$ were used to estimate the electron excitation temperature for these transitions. Results provide only a rough idea of the electron energies because these estimates assume a Maxwellian distribution of electron energy, excitation from the ground state, and purely radiative emission, assumptions that do not hold in moderate pressure RF discharges. Argon line spectra were used to estimate the concentrations of metastable Ar excited states using the branching method described in detail below. [28, 29,30]

## II.3.a. Branching Method for estimating concentrations of metastable states.

Information about the concentrations of metastable Ar* can illuminate the role of metastable gases in $CH_4$ decomposition, such as by the suggested energy pooling and subsequent penning dissociation and dissociative ionization.[31] Concentrations of metastable Ar* states can be assessed using tunable diode laser absorption or laser-induced fluorescence,[18] but the use of laser-based methods is not always practical owing to limited optical access or plasma instabilities that complicate long signal accumulations. Here, we used OES and Ar line ratios to deduce information regarding Ar*. The theory of the method was developed by M Schulze (2008) and uses the idea that the photons emitted due to the transitions from common upper states to resonant or metastable lower states can be reabsorbed.[30] The self-absorption of these photons will affect the observed line intensities such that the ratios of lines intensities of these transitions will be sensitive to the concentrations of the metastable or resonant states. The concept of escape factors is used to write a set of nonlinear equations for the line ratios, which do not depend on the densities of the upper states. The rate of photons seen by the detector can be written as[29]:

$$I_{ij} = c\gamma_{ij}(n_j)A_{ij}n_i,  \quad (1)$$

where $I_{ij}$ is the line-of-sight averaged spectral line irradiance from a given observation column, $c$ is the geometrical factor for this column, $A_{ij}$ is the Einstein coefficient, $\gamma_{ij}(n_j)$ is the escape factor[32,33] dependent on $n_j$ is the concentration of the lower state (1s), $n_i$ is the concentration of the upper states (2p). Both $n_j$ and $n_i$ are unknown. For transitions from the same upper state $i$ (2p) to different lower states $j$ and $k$, the ratio $\frac{I_{ij}}{I_{ik}}$ is independent of $n_i$, the concentration of the upper states, and the geometrical factor c. We can re-write it as:

$$\frac{I_{ij}A_{ik}}{I_{ik}A_{ij}} - \frac{\gamma_{ij}(n_j)}{\gamma_{ik}(n_k)} = 0 \quad (2)$$

The escape factor $\gamma_{ij}(n_j)$, represents the fraction of emitted photons that escape reabsorption to be detected by the optical detector. It decreases when $n_j$ increases because more emitted photons are reabsorbed. Reabsorption is significant even in optically thin plasma owing to the long lifetime of the metastable states. Many simplified expressions are used to estimate the escape factor.[34,35] The estimation below is based on line-of-sight light collection and assumes a uniform distribution of absorbers and emitters. The assumption of uniformity does not hold in many RF discharges, but it does provide an estimation that returns meaningful trends when compared to LIF results.[29]

We can approximately calculate $\gamma_{ij}(n_j)$[29] as $\gamma_{ij} \approx \frac{1}{1+k_{ij}l}$, where $l$ is the plasma size and $k_{ij}(n_j)$ is the reabsorption coefficient.

$$k_{ij}(n_j) = \frac{\lambda_{ij}^3}{8\pi^{3/2}} \frac{g_i}{g_j} A_{ij} n_j \sqrt{\frac{M}{2k_B T_g}} \quad (3)$$

Where $\lambda_{ij}$ is the emitted wavelength; $g_i$ and $g_i$ are the statistical weights of the upper and lower states, respectively; M is the atomic mass; $k_B$ is the Boltzmann constant; $T_g$ is the gas temperature; $A_{ij}$- Einstein coefficient. The $\gamma_{ij}$ is calculated for each line, and $n_j$ can be calculated using the least-squares method to minimize the following equation: [30,29]

$$\sum_{m=1}^{5}\left[\left(\frac{I_{ij}A_{ik}}{I_{ik}A_{ij}}\right)_m - \left(\frac{\gamma_{ij}}{\gamma_{ik}}\right)\right]^2 = 0 \quad (4)$$

The line wavelengths, corresponding transitions, and Einstein coefficients are listed in Supplementary Material, Table S1.[28] Note that the 750.39 nm line was not used due to a significant overlap between this line and the neighboring 751.54 nm Ar line, and the 747.12 nm line was not used because it was too weak. Note that the collision cross-sections are not included in this procedure, and no assumptions regarding the EEDF are necessary. On the negative side, the method is subject to intensity fluctuations due to plasma instabilities or inhomogeneities and works well only if reabsorption is fairly significant, but the plasma is not opaque, so on the boundary between optically thin and optically thick plasma. [29,35] In the future, we are planning on using LIF for the measurement of the Ar $1s^5$ concentration in this reactor.

**III. Results:**
**III.1. Electrical measurements**

The voltages and currents measured for several cycles are shown in Figure 2. The capacitance of the quartz tube/electrode system was estimated and used to compute the displacement current and resulting conduction current (Figure 2). In all inert gases alone, the phase angle between the applied voltage and current is low, <30°, and the current is high (~ 200 mA). With the addition of $CH_4$, at the same power setting, the current decreased, and the phase angle increased to 57°. A low phase angle means that resistive power losses are higher than capacitive losses, and the discharge is in the gamma mode.[36–39] The increase in the phase angle with the addition of $CH_4$ is associated with a decrease in the discharge current and an increase in the discharge voltage. This is typical for the addition of molecular gases that deplete the lower-energy electrons owing to vibrational excitation and the higher voltage required for breakdown. [39,40] Similarly, for the same input power, the current is lower and the peak voltage is higher for the molecular gases, $H_2$ and $CH_4$ when used alone without an inert gas. In Figure 2 c, we see that at the same input power (100 W), the current in 100% $CH_4$ was lower than in its mixtures with inert gases. To reach a current comparable to that in a mixture of Ar/5% $CH_4$ the input power needs to be 150 W, or in pure $CH_4$, 200 W to reach 0.26 A in Ar. The increase in the input power of $CH_4$ also led to a substantial increase in the peak applied voltage, with the peak voltage reaching 1060 V at 200 W (Table 1). Plasma density was estimated from the measured current. For example, the current for Ar at 100 W is 0.25 A. In similar discharges, the measured and/or simulated plasma density was ~ $10^{17}$-$10^{18}$ $m^{-3}$. [34,38,39]

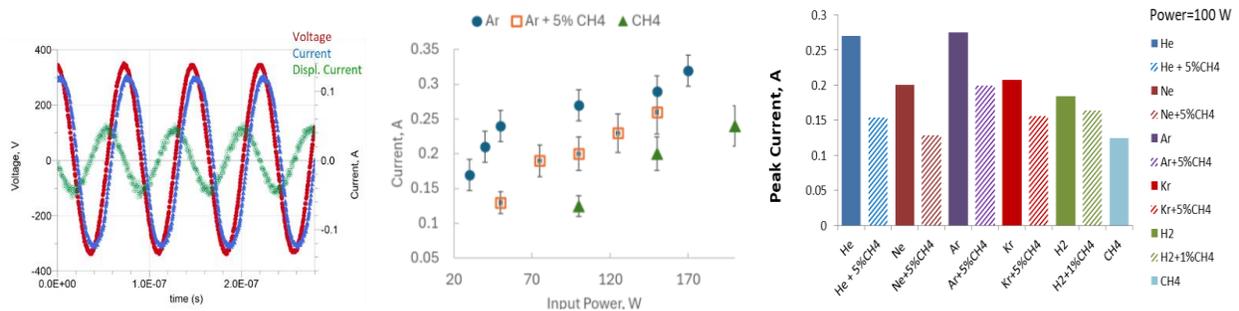

**Figure 2.** (a) Typical voltage and current traces for a discharge in Ar gas showing a mostly resistive current in phase with the voltage and the calculated displacement current (using estimated system capacitance ~1.2 pF). (b) Current as a function of the input power as set on the power source for discharges in Ar, Ar+ 5% CH4, and 100 % $CH_4$. The pressure and flow rates were kept constant at 50 sccm total flow rate and 2.5 – 2.7 torr. (c) Average peak current for the discharge at 2.6 torr and 100 W in $CH_4$, the inert gases, He, Ne, Ar, Kr and their mixtures with $CH_4$. The experimental error is within 5% and therefore the observed differences are statistically significant.

Table 1: Electrical measurements for Ar, Ar/$CH_4$ and for $CH_4$ at 100 W, 150 W, and 200 W input power.

| GAS | Input Power, W | Peak Voltage measured at the electrode, V | Peak Current, A |
|---|---|---|---|
| Ar | 100 | 660 | 0.26 |
| Ar+5%CH$_4$ | 100 | 760 | 0.20 |
| CH$_4$ | 100 | 800 | 0.13 |
| CH$_4$ | 150 | 920 | 0.20 |
| CH$_4$ | 200 | 1060 | 0.24 |

**III.2. OES: CH$_4$ and mixtures with Ar, He, Ne, Kr**

In this section, we report the optical emission spectroscopy (OES) results. Survey spectra were used to track the relative intensities of the characteristic atomic lines and molecular band emissions, evaluate the vibrational and rotational excitation temperatures, estimate the electron excitation temperature, and assess the changes in metastable concentration with the addition of methane.

All the spectra shown below (Figures 3-5) were taken at the same pressure, $2.5 \pm 0.1$ torr, and a flow rate of 50 sccm. The input concentrations of CH$_4$, background gas, and input power were varied during the experiments. The spectra of the discharges with added CH$_4$ in the 350 – 700 nm range have emission bands attributed to CH (389 nm, 430 nm), C$_2$ (516 nm, 470 nm, etc.), H$_2$ (581.5 nm), line emission H$_\alpha$ (656 nm), and under some conditions, H$_\beta$ (Figure 3). We tracked the relative intensities of these bands as a function of the input power and the proportion of CH$_4$ in the Ar background gas (Figures 3 and 4). Interest in this emission region is motivated by the reported correlations between the relative intensities of these bands and lines with the gas and solid products in plasma deposition and decomposition experiments involving methane and other hydrocarbons.[41–43]

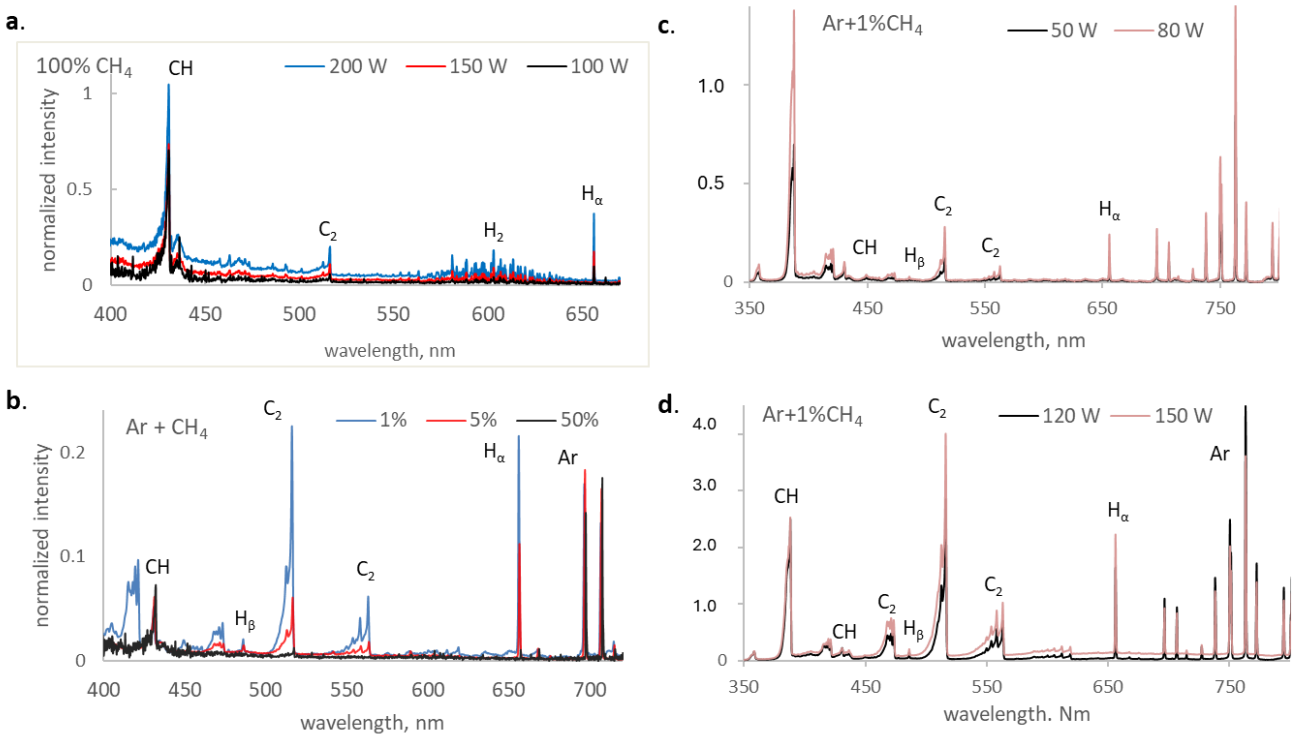

**Figure 3.** Optical emission spectra for RF 13.56 MHz discharge in $CH_4$ and Ar/ $CH_4$ at 2.6 torr and 50 sccm total flow rate in the 350 nm – 700 nm range. (a) CH, $C_2$, $H_2$, and H emission bands in 100% $CH_4$ increase in intensity as the input power increases with the $C_2$ and $H_2$ becoming prominent for the input power > 150 W. (b) Spectra for the discharge in Ar/$CH_4$ and the input power of 100 W show a significant increase in the intensity of the $C_2$, band emissions with decreasing concentration of $CH_4$. (c, d) Relative emission of the bands and lines in the 350 – 700 nm strongly depends on the input power. (c) In the lower power range, 50 W – 100 W, CH radical emission bands at 389 nm and 430 nm dominate the emission spectrum with an overall increase with increasing power. (d) When the power exceeds ~120 W, the relative band intensities change, with a sharp increase in the $C_2$ swan bands (516 nm, 470 nm, etc.), which dominate the spectrum at the higher power.

For 100% $CH_4$ under our experimental conditions, the strongest emission was attributed to CH radical (Figure 3a). At 50 W, the $C_2$ and $H_2$ bands and $H_\alpha$ line were very weak or undetectable. As the power increased from 50 to 200 W, the intensities of $C_2$ and $H_\alpha$ increased. The most drastic differences in $C_2$, H, and $H_2$ emissions were observed at ~200 W (Figure 3a).

When $CH_4$ is added to the background Ar gas, a higher concentration of $CH_4$ leads to a decrease in the overall emission as the discharge dims (note also the lower current at the same input power). At lower proportions of $CH_4$ in Ar, < 5%), we observed an increase in the relative intensity of the $C_2$ band with decreasing proportion of $CH_4$, possibly due to a higher decomposition (Figure 3 b).

An interesting trend was observed when the input power was increased for the same 1% $CH_4$ in Ar (Figure 3 c, d). At lower power, from discharge initiation until ~ 120 W, the most prominent band in the spectrum at 389 nm (386 – 390 nm) corresponds to the $B^2\Sigma^- \rightarrow X^2\Pi$ CH-band and the increase in power leads to the increase in the band intensities with a stronger increase in the $C_2$ bands and H atomic emissions. A rather abrupt change occurred when the power exceeded 120 W. The relative band intensities change, with a sharp increase in the $C_2$ swan bands (516 nm, 470 nm, etc.), which then dominate the spectrum at higher powers. This transition from a dominated to $C_2$ dominated spectrum is observed when we compute the intensity ratios of $C_2$/CH, $C_2$/H, and H/CH as a function of the input power (Figure 4). The intensity ratio of the $H_\alpha$ line increased monotonically with power, but the $C_2$ emission increased abruptly in the 100 – 120 W range. The sudden increase in $C_2$ emissions may indicate a mode transition during discharge. As mentioned above, it is usually associated with a higher degree of decomposition and correlates with a shift to more $C_2H_x$ gas products and carbon solids.[44] Our experimental results on gas products and solids are presented in the FTIR and solid product sections below.

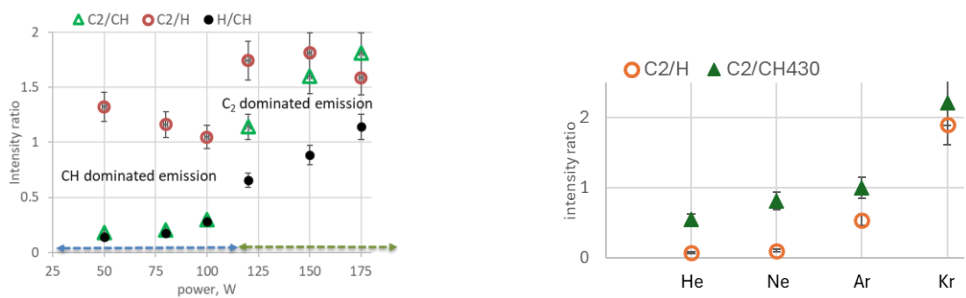

**Figure 4.** (a) The $C_2$/CH, $C_2$/H, and H/CH peak intensity ratios from the emission spectrum in the 350 – 700 nm range as a function of power. In the 100 – 120 W power region, $C_2$/CH, the $C_2$/H ratios increase abruptly, indicating a plasma regime change. (Note: The markers for the data points for H/CH and $C_2$/H overlap at 50, 80, and 100 W.) (b) The $C_2$/CH, $C_2$/H, and H/CH peak intensity ratios from the emission spectrum in the 350 – 700 nm range for different inert gas mixtures with 5% CH4. The total flow rate was 50 sccm, and the input power 100 W.

The optical emission range of 350 – 700 nm was also explored for combinations of methane with other inert gases, He, Ne, and Kr, in addition to Ar. results summarized in Figure 5 were obtained while controlling all other external experimental parameters: input power, pressure, total flow rate, and $CH_4$ concentration: 100 W, 2.5 torr, 50 sccm, and 5% of the total mass flow, respectively. The observed spectral features changed with different background gases. For example, the atomic H emission was strongest in He and Ne and lowest in Kr. The relative strength of the $C_2$ band emission increased in the order He < Ne < Ar < Kr (Figure 4 b). Similar to the changes in the $C_2$ band emission with power, the relative increase in the $C_2$ emission intensity may be correlated with the degree of methane decomposition, gas products, and tendency to form crystalline carbon products. [12,45,46] We present these correlations in Section III.5.

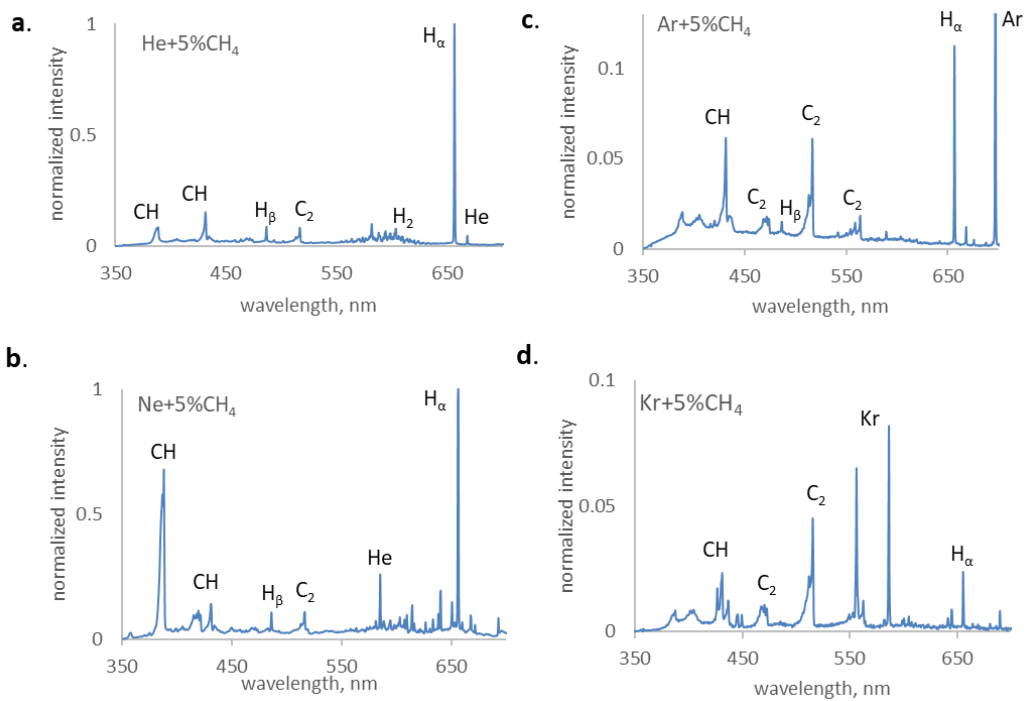

**Figure 5.** Optical emission spectra for 5% $CH_4$ in He, Ne, Ar, and Kr. The spectra are normalized to the highest intensity in the 300 – 800 nm range, which are the atomic Ar and Kr lines respectively in the Ar/CH4 and Kr/CH4 mixtures. (a) For 5% $CH_4$ in He, the highest intensity line is $H_\alpha$ and the spectrum contains the emission bands attributed to CH, $C_2$, $H_2$, and $H_\beta$ line. (b) In Ne, we see a significant increase in the CH 389 nm emission band but the $H_\alpha$ line remains the strongest. (c) In 5% $CH_4$ in Ar, bthere is a notable increase in the $C_2$ swan band emission. (d) Similarly, inKr + 5% $CH_4$ there is an increase in the relative intensity of the $C_2$, band emissions and a decrease in the $H_\alpha$ emission.(13.56 MHz, 100 W discharge in at 2.6 torr and 50 sccm total flow rate).

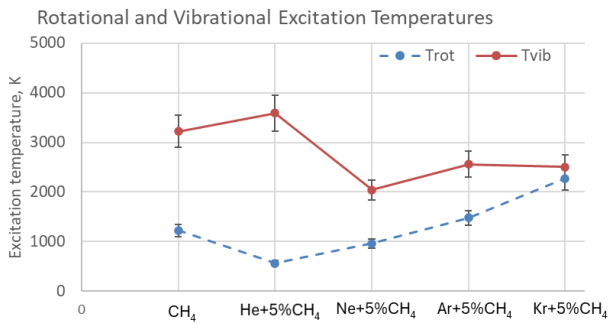

**Figure 6.** Rotational and vibrational excitation temperature as a function of the background gas for 100 % $CH_4$, and 5% $CH_4$ in He, Ne, Ar, and Kr. The discharge was excited by a 13.56 MHz RF power source, 100 W input power, at a pressure of 2.5 torr, and a total flow rate of 50 sccm. Tvib and Trot were determined by spectral fitting of the $C_2$ 516 nm ro-vibrational band using OES-Toolbox open-source software. The light was collected from a discharge region between the electrodes.

For methane and its mixtures with inert gases, $C_2$ swan bands at 516 nm were fitted using OES-Toolbox open source software.[26] The software uses a collisional radiative model to generate a synthetic spectrum and fit it to the experimental spectrum with the rotational and vibrational excitation temperatures as the fitting parameters. The rotational excitation temperature $T_{rot}$ was assumed to be reasonably close to the gas translational temperature under our experimental conditions. The non-equilibrium between electronically excited vibrational states and rotational and translational collisional excitations exists in non-equilibrium plasma due to the much longer vibration-translation relaxation than rotation-tranlational relaxation.[1] Representative fitted spectra and the resulting values for $T_{rot}$ and $T_{vib}$ are shown in the Supplementary Material (Figure S1). In 100% $CH_4$, $T_{rot} \approx T_{gas}$ increased with power from 900±100 K at 150 W to 1200±100 K ($T_{vib}$ = 4800±150 K) at 200 W. The uncertainty is determined statistically from measurements of the same operational parameters and indicates the degree of irreproducibility of the plasma conditions. $T_{rot}$ and $T_{vib}$ were determined for the same experimental conditions: 100 W input power, 2.5 torr, and a total flow rate of 50 sccm. The plot shows that $T_{rot}$ and $T_{vib}$ approach each other, with the ratio $\frac{T_{vib}}{T_{rot}} \approx 6$ for He, then dropping to $\approx 2$ for Ne, $\approx 1.7$ for Ar, and $\approx 1$ for Kr. Therefore, the discharge tends to approach thermal equilibrium with the change in the background gas, changing from highly non-equilibrium to approaching "warm" plasma conditions for Kr. These trends are important for tuning the product selectivity. [3,8,12]

The ratio of the line intensities, $\frac{I_{ji}}{I_{kl}}$, for the hydrogen $H_\alpha$ and $H_\beta$ lines was used to estimate the electron temperature. Assuming a Maxwellian distribution of excited states, optical transparency, excitation from ground states, and radiative de-excitation, the line intensity ratio exponentially depends on the difference between the excited states:

---

[1] Here Tgas represents the translational degrees of freedom

$$\frac{I_{ji}}{I_{kl}} = \frac{\lambda_{kl}A_{ij}g_j}{\lambda_{ij}A_{kl}g_k} e^{\left(\frac{E_k-E_j}{k_B T_{ex}}\right)}, \tag{5}$$

Where $\lambda_{kl}$ and $\lambda_{kl}$ are the wavelengths of the lines, 656.3 nm and 486.1 nm respectively; $A_{kl}$, $g_k$ are the Einstein coefficient and statistical weight of the upper states k and j and $E_k$ and $E_j$ are the energy of these states. Equation (5) is solved for $T_{ex}$ given the measured line intensities.[47] The temperature is in the range of 0.8 – 1 eV across most of our experimental conditions, which agrees with the values reported for the RF CCP in our pressure range.[36,48] This method has been used in many plasma applications, including diamond deposition and plasma decomposition of methane and other hydrocarbons,[49,50] but in hydrocarbon plasma, the method can underestimate the electron temperature because excited hydrogen H* is produced in decomposition reactions such as for example, $e + CH_3 \rightarrow CH_2 + H^*$.[51] In addition, the plasma density in our study may be too low for the above assumptions (<$10^{13}$ cm$^{-3}$) particularly the non-Maxwellian distribution of the excited states.

Because Penning dissociation and stepwise excitation have been implicated as the main reasons inert gases have a strong effect on the discharge and methane decomposition, we used OES analysis to estimate the Ar metastable densities using the branching method.[31,45,52,53] Figure 7a shows the Ar emission lines used in the branching method calculations of $Ar_m^*$ metastable states. The method returned the concentrations of the $1s^5$, $1s^3$ metastable, and $1s^4$ resonant states. The concentrations were of the order of ~$10^{11}$ cm$^{-3}$ without the addition of methane and decreased by almost an order of magnitude for 5% CH$_4$ in Ar. The observed decrease in metastable concentrations is expected because of the very fast quenching pathways of molecular gases (CH$_4$, H$_2$, N$_2$, etc.). The densities and trends reported here agree with the experimental and theoretical values in the literature, but we plan to verify our observations in the future using laser-induced fluorescence.[18,54,55] The measurements for other inert gases, such as He, Ne, and Kr, are beyond the scope of this study.

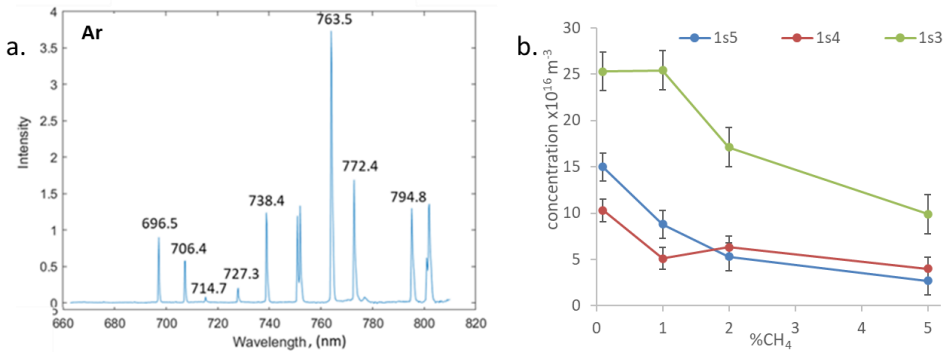

**Figure 7.** (a) Ar emission lines used in the branching method and (b) concentrations of the metastable, $1s^5$ and $1s^3$, and the resonant state $1s^4$ for different percentage of CH$_4$ in Ar. The concentration of $1s^2$ was not found due to the lack of intensity or resolution for the required emission lines. RF 13.56 nMHz, 100 W, 50 sccm, 2.5 torr.

**III.3. FTIR measurements of gas products and CH$_4$ decomposition.**

FTIR absorption spectra were fitted using the HITRAN database and a homemade Python routine to determine the concentration of $C_2H_x$ and $CH_x$ products.[56] The % decomposition was calculated as $\%decomposition = \frac{100*(CH_{4initial}-CH_{4plasma})}{CH_{4initial}}$, where $CH_{4initial}$ and $CH_{4plasma}$ are concentrations in ppm. For 100% $CH_4$, decomposition increased from 30% at 100 W to 74% at 200 an input power. This correlates with the increase in the intensity of the decomposition fragments in OES, as shown above. In mixtures with Ar, the degree of decomposition depended on the % of $CH_4$ added to Ar (Figure 8), reaching 99.5% for <5% initial concentration of $CH_4$. The main gas product under these conditions was $C_2H_2$. The addition of He, Ne, and Kr resulted in different decomposition rates with Kr at 98% approaching Ar, and Ne only reached 69% at the same input power, total flow rate, and pressure. He has a high degree of decomposition of 90% but as we show below, no soot formation.

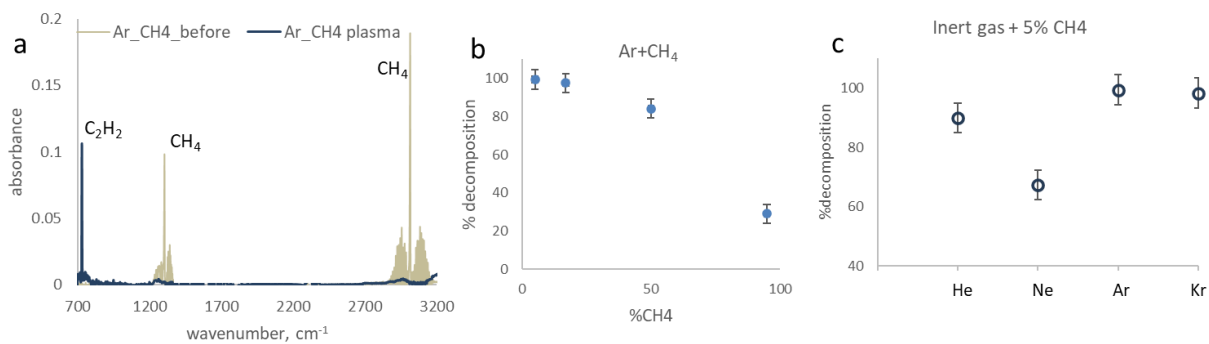

**Figure 8** (a) A representative FTIR spectrum showing the formation of $C_2H_2$ in the discharge of Ar+5%$CH_4$. (b) The degree of decomposition decreases with increasing concentration of methane at the same input power. (c) The decomposition is the highest for Ar and Kr, both showing a nearly complete destruction of $CH_4$, but the addition of He also leads to ~90% decomposition. The input power was 100 W, the total flow rate, 50 sccm, and the pressure 2.5 torr for all these results.

**III.4 Fast frame imaging.**

The imaged region of the quartz tube between the Cu-band electrodes is shown in Figure 9a. The images of this region in the discharges revealed several distinct plasma modes at different power and/or gas combinations. The discharge in 100 % $CH_4$, 100% He, and 100% Ne (Figure 9b, c, d) is in diffuse mode under our operating conditions and appears uniform even at 10000 fps, the fastest available frame rate with Phantom-4. In contrast, the discharge in 100% Ar and 100% Kr was striated. Striations are not uncommon in DC and RF discharges at low and moderate pressures but are not well understood. The speed of the moving striations was estimated from the analysis of fast frame images taken at 10000 fps and 10 μs exposure. The striations here move at speeds of ~ 5 m/s, which is comparable to the speed of the gas flow in the tube (see Discussion for details of the gas speed calculation). The wavelength of the striations was approximately 6 mm in Ar and 7 mm in Kr (Figure 9 g, i). Striations are strongly affected by the mean free path; hence, the difference between the wavelengths in Ar and Kr may be due to the measured difference in the temperature, which leads to a difference in the gas density of the two gases.

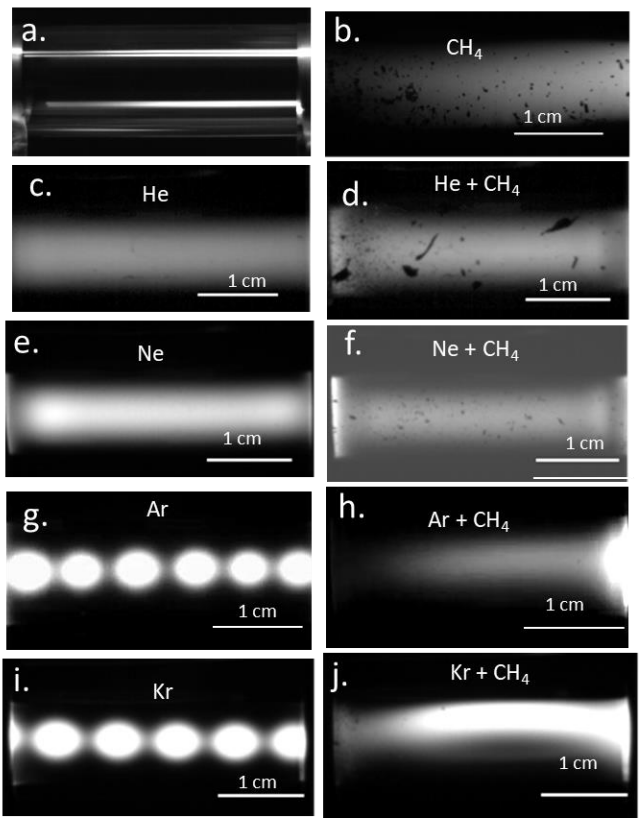

**Figure 9.** Fast frame images of the discharge in the individual gases and inert gas/methane mixtures. (a) For reference purposes, the copper band electrodes are seen on both sides of the imaged interelectrode part of the quartz tube. Discharge is in the diffuse mode in 100% $CH_4$, He, and Ne (b, c, and e) and in the mixtures of He+5%$CH_4$ and He+5%$CH_4$ (d, f). Deposits are visible on the tube walls during the discharge in 100% $CH_4$ and He+5%$CH_4$ and He+5%$CH_4$. In contrast, moving striations occur in Ar and Kr (g, i) with the wavelength slightly longer in Kr (7 mm) than in Ar (6 mm). The tube walls remain clear in mixtures Ar+5%$CH_4$ and Kr+5%$CH_4$ (h, j), and the discharge filament moves azimuthally. The videos of the moving striations and filaments are found in the Supplementary Information section. Input power 100 W (150 W for 100% CH4), 50 sccm total flow rate, 2.5 torr.

Copious deposits on the tube walls were observed in 100% $CH_4$ and in $CH_4$ mixtures with He and Ne (Figure 9 b, d, f). These deposits appeared within <1 min of operation, and the discharge remained diffuse. The deposits were collected and tested, as reported in the section on solid products. Under similar experimental conditions in Ar+5%$CH_4$ and Kr+5%$CH_4$ (Figure 9h, j), the tube remains clear, the discharge contracts, and the resulting filament moves somewhat erratically azimuthally inside the ground electrode. In Section V, we discuss the nature of this unstable behavior and its possible relationship with the products. Fast-frame videos are available in the Supplementary Information section. The onset of the contraction into a filament and the movement of the filament are dependent on the input power and exhibit hysteresis continuing at a lower power when the power is increased than when it is originally increased. For example, in Ar+5%$CH_4$ the contraction instability does not appear until the input power >150 W, but the unstable behavior continues to <100 W. This behavior was similar for Kr+5%$CH_4$.

Although the walls of the tube remained clean during the 20 – 30 min long experiments, soot-like deposits were collected on the filter placed in the gas line. Therefore, it appears that in Ar and Kr, with a small addition of $CH_4$ solid material is produced in the gas phase of the discharge under our experimental conditions. The characterization of the solid material is discussed in the next section (Section III.5).

**III.5. Solid product analysis.**

Two types of solid deposits were collected and analyzed in this study, each corresponding to a different discharge mode. When the discharge was in the diffuse mode, deposits appeared on the tube walls, and no deposits were collected on the filter (Figure 10 a, b). In contrast, the contracted unstable discharge mode produced black powder collected on the filter, while the tube walls remained relatively clean. The position of the inflow filter is shown in Figure 1.

For example, in 100% $CH_4$ at 200 W, a substantial deposit was observed on the tube walls, mostly in the electrode region, with some additional deposits between the electrodes. No visible or measurable deposits were collected from the filters. Visually, the solid deposits consist of macroscopic flakes, particles, and glassy deposits of various shades of brown and black color. The SEM images show amorphous micrometer-sized layers and glassy amorphous deposits (Figure 10 c, d). TEM imaging was not possible because the glassy deposits could not be properly dispersed for the TEM analysis.

In cases of the contracted moving filaments in Ar+5%$CH_4$ and Kr+5%$CH_4$, the filter was covered with black powder (Figure 10 b). The deposits collected from the filter were imaged using TEM and high-resolution TEM (HRTEM). The material appears graphitized with layers 2-3 nm thick layers (measured using ImageJ®  software). The layer curves often form circular structures, as shown in Figure 10e and f. Additional studies are in progress to address the statistical frequency of specific structures as well as Raman spectra analysis to assess the degree of crystallinity.

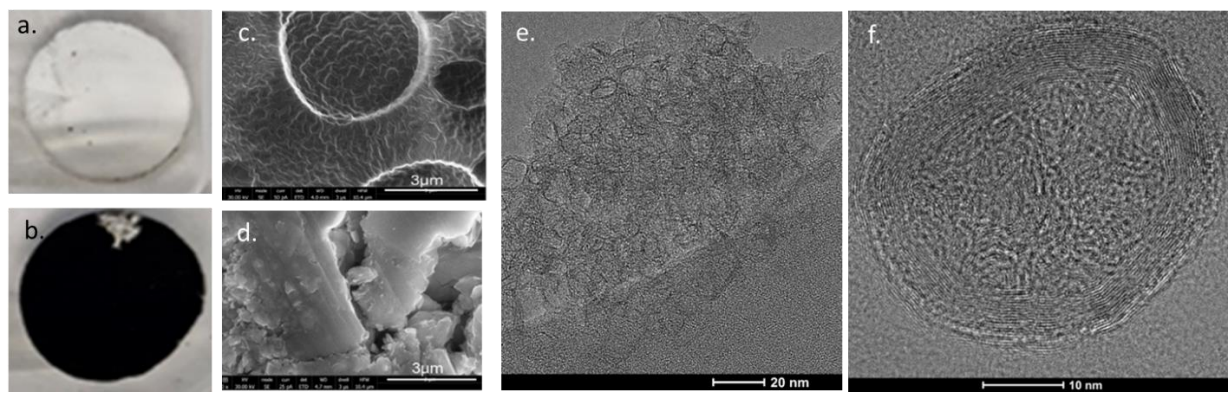

**Figure 10** Sample solid products, deposits on the walls and on the filter. (a) Filter inserted in the gas line has no deposit for diffuse discharge in 100% $CH_4$ (200 W) and for He+5%$CH_4$ and Ne+5%$CH_4$ (100 W). (b) For discharges in Ar+5%$CH_4$ and Kr+5%$CH_4$, the filter is covered with black powder. (c, d) SEM images of the wall deposits show amorphous glassy deposits (c) and µm size amorphous layers. (e) Typical TEM

image of graphitized deposits on the filter (b) showing graphitic layers and a sample circular structure (f). The discharge conditions: 50 sccm flow rate, 2.6 torr, 200 W.

### III. 6. Summary of the Experimental Results.

In this work, we conducted electrical, OES, FTIR, and solid product measurements and analysis for RF 13.56 MHz discharge in a quartz tube with external Cu electrodes, 10 mm wide, 35 mm apart, for a total distance of 55 cm. The pressure was 2.4- 2.6 torr, and the total flow rate was 50 sccm for all gases and gas combinations including 100% $CH_4$, Ar, He, Ne, Kr and the mixtures, of Ar with 0.5%, 1%, 2%, and 5% $CH_4$, and He, Ne, and Kr with 5% $CH_4$.

The discharge in all gases in the range of 50 – 200 W appeared to be in gamma mode, as determined by the current amplitude and the phase difference between the current and voltage. We have only order-of-magnitude estimates of the plasma density and electron excitation temperature so far, $n_e$ ~$10^{12}m^{-3}$ and $T_{exc}$~0.6 – 1 eV. The discharge is accompanied by substantial gas heating and results in methane decomposition and the formation of different gas and solid products depending on the discharge mode (diffuse or contracted). The results are summarized in Table 2.

**Table 2.** Summary of the results for 5% $CH_4$ mixtures with inert gases for the pressure of 2.5 – 2.7 torr, 50 sccm total gas flow, 150 W input power.

| Gas | Instabilities 95% + 5% $CH_4$ | Discharge mode | Trot, K | % decomposition at 150 W | Product: solid/gas |
|---|---|---|---|---|---|
| $CH_4$ | no | diffuse | 1000 K | 67% | Amorphous/$C_2H_x$ |
| He | no | diffuse | 600 K | 90% | Amorphous |
| Ne | no | diffuse | 960 K | 67% | amorphous |
| Ar | Yes | contracted | 1500 K | 99.5% | Graphitized carbon, $C_2H_2$ |
| Kr | Yes | contracted | 2200 K | 98.4% | Graphitized carbon, $C_2H_2$ |

### V. Discussion:

Our experiments revealed the correlations between the discharge mode and behavior, methane decomposition, and product selectivity. To explain these observations, we first address the observed discharge behavior, including the diffuse and contracted modes, and then relate this behavior to product selectivity.

The addition of even <1% of $CH_4$ to an RF Ar discharge has an immediate effect on the discharge. The presence of $CH_4$ introduces efficient inelastic electron collision channels, particularly vibrational excitation, which strongly depletes the electron energy-distribution function. Consequently, the sustaining voltage increases, the electron density decreases, and the overall plasma conductivity is reduced. Experimentally, the $CH_4$ addition produces a marked reduction in discharge current, dimming of the positive column, and requires higher power. In Ar, striations vanish at a certain concentration of $CH_4$ (depending on the input power), and below ~100–150 W, the discharge becomes diffuse with a reduced current compared to pure Ar (Figure 2, Table 1). The reduction of Ar metastable concentration with increasing $CH_4$ confirms strong quenching of metastables.[64]

These admixture effects set the stage of plasma contraction. Under the operational conditions of this study, the discharge exists either in a diffuse or in a contracted mode, but the contracted mode recorded by fast frame imaging appears in Ar and Kr with the addition of small amount <5% of $CH_4$ at input power >100 W. Contracted mode is not observed at low voltages and does not appear in He or Ne even at input power ~200 W.

We now provide a qualitative explanation of the contraction in Ar with an admixture of $CH_4$.

**V.1. A simple model to understand plasma contraction**

The contraction recorded in our RF flow reactor may be an ionization thermal instability that occurs in weakly ionized, non-equilibrium, longitudinal plasma flows in a tube. Two conditions must be met for this type of contraction to occur.[57] First, the ionization frequency should be higher near the axis of the flow than in the periphery. Our results for the gas temperature from Ar/CH4 OES ($C_2$ emission) from the tube axial region are much higher (1200 – 2000 K) than that of the tube wall at ~450 K. Consequently, the reduced electric field ($E/N$) is higher in the emission region; hence, the condition for a higher ionization frequency in the axial region of the tube is satisfied. With a flow rate of 50 sccm in our experiments, the flow velocities were < 10 m/s; therefore, turbulence and additional heating in the boundary layer near the tube wall can be ignored.

In the second condition, electrons must recombine close to where they are produced, so that they do not have enough time to diffuse to the tube walls from the hot axial region. We discuss this in more detail below but provide a quick estimate here to illustrate that this may indeed be the case when molecular ions are present. Using $D_a \approx 2$ cm² s$^{-1}$ for Ar at 2.5 Torr with $T_e$ ~ 1 eV,[57] the ambipolar diffusion time across the tube (R = 5 mm) is ~20 ms. In comparison, radiative recombination alone would yield electron lifetimes on the order of 10 s, far too slow to counteract the diffusion and localize the plasma. However, with molecular ions formed in the presence of $CH_4$ ($CH_4^+$, $CH_3^+$, $C_2H_2^+$), dissociative recombination rates (β ~ 10⁻⁷ cm³ s⁻¹)[57] shorten electron lifetimes to ~0.1 ms at $n_e$ ~ 10¹¹ cm⁻³ and ambipolar diffusion is then too slow to transport the electrons to the walls.[66,38]

Now, we will explore in more detail the possibility of contraction, the plasma conditions needed for the formation of an axial contracted region by setting up a simple 0-dimensional model. The intent of the model is limited to establishing the critical plasma conditions that enable contraction. A full description of the contracted state itself is beyond the scope of this study.

As stated above, the presence of molecular ions in the plasma is essential for contraction because of the high cross section for dissociative recombination. Although a variety of molecular ions are formed during methane decomposition, it is sufficient to include a few ions to satisfy the conditions for contraction. As more methane continues to decompose, molecular ions are added to the mix, and the contraction increases. Hence, for simplicity, we include in the model the molecular cluster ions, such as $Ar_2^+$, and two types of molecular ions, $CH_4^+$ and $CH_3^+$ which are formed in the first step reactions of $CH_4$ decomposition.

$CH_3^+$ are produced mainly by the $Ar^+ - CH_4$ charge transfer: $Ar^+ + CH_4 \rightarrow CH_3^+ + H + Ar + 1.46\ eV$ with the reaction constant $k_{ch\_tr} = 0.93 \cdot 10^{-15}$ m³/s. $CH_3^+$ represents at least 85% of the produced ions.[58]

In argon plasma, in addition to atomic Ar+ ions formed by direct and step ionization, cluster diatomic $Ar_2^+$ ions are formed as a result of the conversion of atomic ions in triple collisions $Ar^+ + Ar + Ar \rightarrow Ar_2^+ + Ar$ with a reaction rate $k_{conv} = 2.5 \cdot 10^{-43} \left(\frac{300}{T[K]}\right)^{0.75}$ m⁶/s.[57, 59]

The dissociative recombination coefficient of ions $Ar_2^+$ [60]: $\beta_{rec,1} = 9.1 \cdot 10^{-13}(300/T_e[K])^{0.61}$ m³/s. The data on the dissociative recombination constants of $CH_4^+$ and $CH_3^+$ ions, – $\beta_{rec,2}$ and $\beta_{rec,3}$, are scattered. We use the values in [61]: $\beta_{rec,2} = \beta_{rec,3} = 3.5 \cdot 10^{-13}(300/T_e[K])^{0.5}$ m³/s. For the qualitative analysis, we will not consider the numerous subsequent decay products of the molecular ions $CH_4^+$ and $CH_3^+$ but will include the recombination of atomic ions, $Ar^+$ including the three-body collisional recombination and photorecombination [57]:

$$\beta_{rec,4} = 8.75 \cdot 10^{-39} T_e^{-4.5} n_e + 2.7 \cdot 10^{-19} T_e^{-0.75}\ [m^3/s], \tag{6}$$

where $T_e$ in eV and $n_e$ in m⁻³.

The frequency of ambipolar loss to the walls is given by the equation: $\nu_{amb} = D_a/\Lambda_D^2$, where $D_{amb} = \mu_i T_e$ is the ambipolar diffusion coefficient and $\Lambda_D = R/2.4$ is the characteristic radial diffusion length for a cylindrical tube of radius R.[57] For simplicity, assume that the ambipolar diffusion coefficient of the mixture is determined by the argon ion mobility $\mu_i = 0.128(T/300)/p$ m² V⁻¹ s⁻¹ satisfying the values of mobility given in [62]. The rate coefficients of ionization of the $Ar$ and $CH_4$ mixture components $k_{i,Ar}$ and $k_{i,CH_4}$, the excitation rates of the argon Ar* metastable states, $k_{ex,met}$, and the electron temperature $T_e = \frac{2}{3}\varepsilon$ were calculated using the software Bolsig+ [63] for given values of the reduced field amplitude $E_a/N$, its frequency, $\omega$, and partial local concentrations of $Ar$ atoms and $CH_4$ molecules. where $\varepsilon$ denotes the average electron energy.

OES branching showed that $Ar^*$metastable concentrations decreased significantly with the addition of methane. Therefore, we include the following reactions of metastable argon states $Ar^*$: the decay of $Ar^*$ in collisions with electrons, $k_{q,1}$, $Ar^* + e \rightarrow Ar + e$[39]; the pair $Ar^* + Ar \rightarrow Ar + Ar$, $k_{q,2}$, and triple collisions, $k_{q,3}$, with argon atoms $Ar^* + Ar + Ar \rightarrow Ar_2 + Ar$ [65] ; pairwise collisions of metastable argon $Ar^*$ with methane molecules $CH_4$ [64]; step ionization, $k_{i,step}$, $Ar^* + e \rightarrow Ar^+ + e + e$[39]; pairwise ionizing collisions (associative ionization) of metastables $Ar^* + Ar^* \rightarrow Ar^+ + Ar + e$ with rates taken from the reference.[65] Some metastable $Ar^*$ diffuses to and is quenched on the tube walls with a characteristic frequency of $v_D^* = D^*/\Lambda_D^2$, where $D^* = 2.4 \cdot 10^{20}/N$ m²/s. Average radiative lifetime of excited argon atoms, $Ar^* \rightarrow Ar + \hbar v$, $\tau_v = 3 \cdot 10^{-8}$s .[39]

**Table 3**. The relevant processes with their rate constants, included in the 0-dimensional model of the conditions for contraction.

| Process | Rate | Dimensions | Reference |
|---|---|---|---|
| $e + Ar \rightarrow Ar^+ + e + e$ | $k_{i,Ar}(E/N, \omega)$ | m³/s | BOLSIG+ [63] |
| $e + CH_4 \rightarrow CH_4^+ + e + e$ | $k_{i,CH_4}(E/N, \omega)$ | m³/s | BOLSIG+ [63] |
| $e + Ar \rightarrow Ar^* + e$ | $k_{ex,met}(E/N, \omega)$ | m³/s | BOLSIG+ [63] |
| $Ar^+ + Ar + Ar \rightarrow Ar_2^+ + Ar$ | $k_{conv} = 2.5 \cdot 10^{-43} \left(\frac{300}{T[K]}\right)^{0.75}$ | m⁶/s | [40, 59] |
| $Ar^* + e \rightarrow Ar^+ + e + e$ | $k_{i,step} = 1.8 \cdot 10^{-13} T_e^{0.61} \exp(-\frac{2.61}{T_e})$ | m³/s $T_e$ [eV] | [39] |
| $Ar^* + Ar^* \rightarrow Ar^+ + Ar + e$ | $k_{i,ass} = 6.2 \cdot 10^{-16}$ | m³/s | [65] |
| $Ar_2^+ + e \rightarrow Ar + Ar$ | $\beta_{rec,1} = 9.1 \cdot 10^{-13} \left(\frac{300}{T[K]}\right)^{0.61}$ | m³/s | [60] |
| $Ar^+ + e + e \rightarrow Ar + e$ $Ar^+ + e \rightarrow Ar + hv$ | $\beta_{rec,4} = 8.75 \cdot 10^{-39} T_e^{-4.5} n_e + 2.7 \cdot 10^{-19} T_e^{-0.75}$ | m³/s $T_e$ [eV]; $n_e$ [m⁻³] | [40] |
| $Ar^* + e \rightarrow Ar + e$ | $k_{q,1} = 3 \cdot 10^{-13} T_e^{0.51}$ | m³/s $T_e$ [eV]; | [39] |
| $Ar^* + Ar \rightarrow Ar + Ar$ | $k_{q,2} = 3 \cdot 10^{-21}$ | m³/s | [65] |
| $Ar^* + Ar + Ar \rightarrow Ar_2 + Ar$ | $k_{q,3} = 1.1 \cdot 10^{-43}$ | m⁶/s | [65] |
| $Ar^* \rightarrow Ar + hv$ | $\tau_v = 3 \cdot 10^{-8}$ | s | [39] |
| $Ar^+ + CH_4 \rightarrow CH_3^+ + H + Ar + 1.46\ eV$ | $k_{ch\_tr} = 0.93 \cdot 10^{-15}$ | m³/s | [58] |
| $Ar^* + CH_4 \rightarrow Ar + neutrals$ | $k_{q,CH_4} = 6 \cdot 10^{-16}$ | m³/s | [64] |
| $CH_4^+ + e \rightarrow neutrals$ | $\beta_{rec,2} = 3.5 \cdot 10^{-13} \left(\frac{300}{T[K]}\right)^{0.5}$ | m³/s | [61] |
| $CH_3^+ + e \rightarrow neutrals$ | $\beta_{rec,3} = 3.5 \cdot 10^{-13} \left(\frac{300}{T[K]}\right)^{0.5}$ | m³/s | [61] |

Calculations were performed under a selected set of experimental conditions: the initial composition of $Ar:CH_4 = 0.95:0.05$, tube radius of $R = 0.5$ cm; average gas temperature in the flow $T = 1000$ K, pressure $p = 2.5$ Torr; and flow velocity $u = 10$ m/s. The ionization and decomposition of methane in the plasma changed the partial composition of the mixture. However, because we are interested in the discharge regime when contraction has not yet developed and the degree of ionization is relatively low, we assumed that the partial composition of the neutral component of the mixture was unchanged. The relevant processes and their rate constants, included in the 0-dimensional model of the conditions for contraction, are summarized in Table 3.

Note that Ref.[66] considered the transition from a diffuse mode of a DC discharge in inert gases to the contracted regime in a tube at pressures of tens of torr in great detail, both experimentally and theoretically. At elevated pressures, in Refs. [66], the dominant mechanism of charged-particle loss is the dissociative recombination of electrons with the cluster ions $Ar_2^+$. In pure argon under our conditions, the discharge occurs in the diffuse mode because the density of the formed $Ar_2^+$ cluster ions is small and three-body recombination is negligible; hence, dissociative recombination plays a minor role compared to ambipolar escape to the tube walls. The addition of a small amount of methane to argon changed the situation significantly; a noticeable number of molecular $CH_4^+$ and $CH_3^+$ ions, among others, appeared in the plasma. Consequently, dissociative recombination became the dominant mechanism of plasma decay, and the volumetric (diffuse) discharge mode was replaced by a contracted mode.

The model presented here is 0-dimensional and depends only on time. However, because gas flows through the tube, the evolution of the plasma over time in a coordinate system that moves with the gas is equivalent to its evolution in space along the flow. Therefore, different distances x from the inlet ring electrode (electrode downstream with coordinate $x = 0$) correspond to different plasma evolution times $t = x/u$, where $u$ is the average gas flow velocity over the tube cross section. If the flow rate increases, we can expect the diffuse discharge region to become more extended and partial contraction to become more pronounced. Longer residence times or slower gas speed would result in the conditions for contraction satisfied closer to the powered electrode.

Figure 11 shows how the ratio of the ambipolar diffusion frequency to the total frequency of electron recombination losses, $\xi = \frac{v_{amb}}{(\sum_i \beta_{rec,i})n_e}$, changes over time, or equivalently, in the plasma flow in different parts of the tube as a function of the reduced field in the tube.

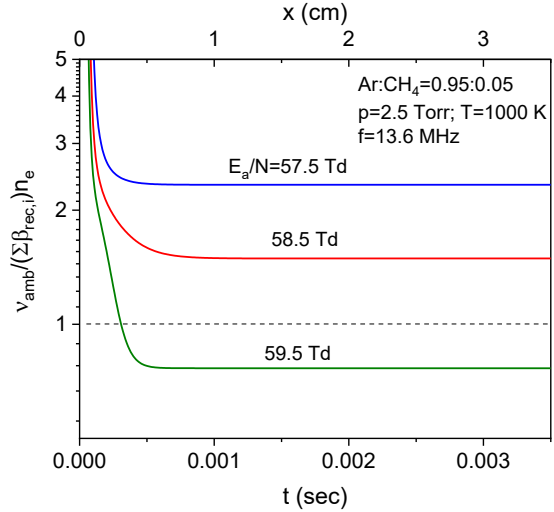

**Figure 11.** The ratio of the frequency of ambipolar plasma escaping the tube walls to the total frequency of recombination losses. The dashed horizontal line corresponds to the boundary between the diffuse (volumetric) plasma of the positive column above the line and the contracted regime below the line.

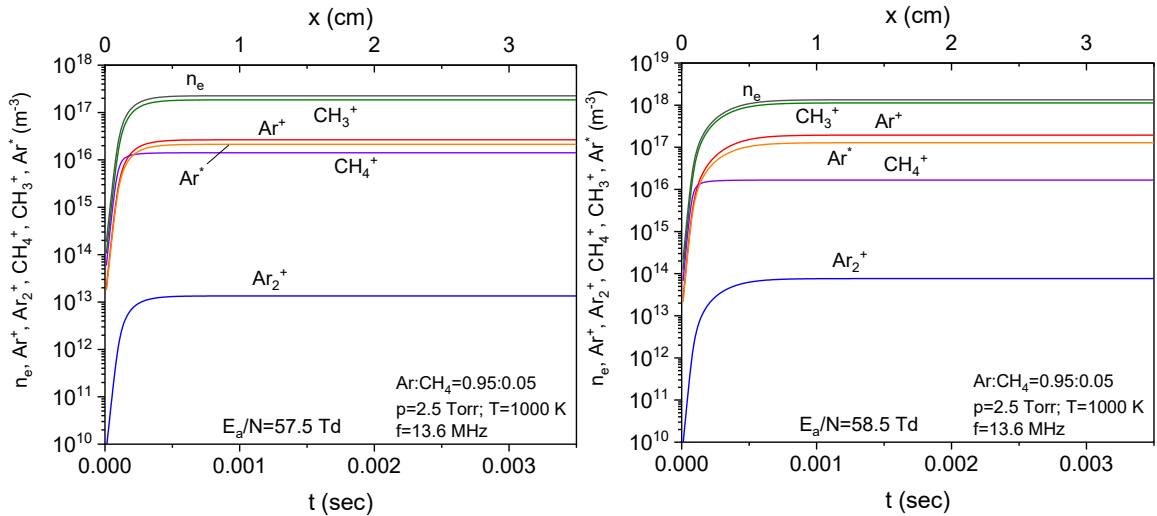

**Figure 12.** The development of plasma in a longitudinal RF capacitive discharge in a longitudinal flow in time and along the tube. (A) diffuse; (B) close to the transition to a contracted regime.

Within the given conditions and amplitudes of the longitudinal electric field in the flow, if $\frac{E_a}{N} < 59$ Td, then according to Figure 11, the calculated values of ξ>1 are valid throughout the tube. Hence, the discharge was sustained in a volumetric diffuse form. As the reduced field increased and $E_a/N$ exceeded 59 Td, the conditions for contraction were satisfied, that is, ξ was less than 1, and the discharge switched to the contracted regime. Figure 12 shows the plasma parameters in the diffuse discharge mode under conditions close to the transition to the contracted state for two values of the reduced field, $\frac{E_a}{N} = 57.5$ and 58.5 Td. The corresponding peak-to-peak estimates of the discharge current amplitudes are $2I_a \approx$ 0.3 and 1.8 A, respectively. The measured peak-to-peak amplitudes of $2I_a \approx 0.5$ A fall within this range.

As previously mentioned, we did not consider the decomposition of methane molecules resulting from pyrolysis in this model. The degree of ionization in the channel and temperature of the gas increased during contraction. This enhances the pyrolysis of the mixture. A thorough examination of these processes requires a self-consistent solution to the problem of RF discharge in gas flow through the tube, considering the external circuit. Such an analysis is beyond the scope of this paper, but it suggests that the presence of molecular $CH_x$ ions is critical for contraction. The interplay between gas heating (high E/N) and the production and recombination of molecular ions drives the contraction.

**V.2. Gas heating for different inert gases**

The importance of gas heating in an RF flow reactor was pointed out in[12] where the authors addressed the possibility of a high-temperature core in such a reactor. In the above discussion, we showed that gas heating, the high-current mode of the discharge, and discharge contraction are interdependent, and that the contracted discharge forms a high-temperature axial filament in the tube. The temperature of this region varies depending on the choice of the inert background gas, with the temperature increasing in the atomic mass order $T_{gHe} < T_{gNe} < T_{gAr} < T_{gKr}$ (Figure 6). To explain the effect of the background gas on heat transfer in the flow reactor, we used simplified estimations based on the energy balance. As mentioned above, full self-consistency is required to understand the complex behavior of the discharge, which is beyond the scope of this study. Here, we limit ourselves to rough estimations to gain qualitative insights.

For these estimations, we consider a simplified heat balance equation:

$$nc_p \frac{dT}{dt} = jE_p - q_{cond} - q_{conv} \tag{7}$$

where T is the gas temperature in K, n is the gas number density at the gas temperature T, $n = \frac{p}{k_B T_g}$, $c_p$ is the heat capacity per molecule of the gas, j is the measured current density, $E_p$ is the plasma electric field in the tube between the electrodes, and $q_{cond}$ and $q_{conv}$ are the heat conduction and convection by the flow, respectively,.[39,40,67] In the steady state, this is reduced to

$$jE_p = q_{cond} - q_{conv} \tag{8}$$

To calculate $E_p$, we used the current measured for each gas (Figure 2c) and the estimated plasma reduced electric field from the contraction model above 60 Td. For contracted discharges, the measured rotational excitation temperatures are ~1000 K, so $n \approx 2.4 \times 10^{22} m^{-3}$ and $E_p \approx 1400 \, V/m$. He and Ne do not contract, so the reduced E field may be lower (higher gas density); however, because the measured voltage is higher for gases reaching lower temperatures, we will keep the same electric field as a rough estimate. The estimate of the heat flux by conduction to the walls in cylindrical coordinates with z as the direction along the gas flow tube and r as the direction across the tube is:[68]

$$Q_{cond} \approx 2\pi RL\lambda \frac{\partial T}{\partial r}\bigg]_{r=R} \tag{9}$$

assuming that the temperature gradient is uniform along the tube axis. where $R$ is the tube radius, L is the tube length, and $\lambda$ is the heat conductivity in $\frac{W}{m \cdot K}$.

For contracted discharges, we can approximate the contraction as a hot filament, similarly to[12] but for a steady-state condition, per unit volume:

$$q_{cond} = \frac{2\lambda\Delta T}{R^2 \ln(R/r_o)} \tag{10}$$

$r_o$ denotes the radius of the contracted filament. $ln(R/r) \approx 2$ The diameter of the filament from the images was approximately 3 mm; therefore, so $ln(R/r_o) \approx 1.2$ was used in the calculations.

If the discharge is diffuse, such as in He and Ne, then we can use the approximation for cooling with a heat source throughout the tube and a parabolic form of the temperature gradient in (9), which results in the following expression per unit volume:

$$q_{cond} = \frac{4\lambda\Delta T}{R^2} \tag{11}$$

Hence, we used (10) for Ar and Kr as the background gas and (11) for He and Ne. We used the heat conductivity at low pressure from[69] to $\lambda$. Is 0.17 W/m for He to 0.0546 W/m for Ne, and 0.0333 W/m for Ar. It is reasonable to assume an even lower conductivity for Kr and use the general 0.02 W/m.[12,69,70]

The addition of $\leq 5\%$ $CH_4$ will affect the conductivity roughly in proportion to the added fraction. Since the conductivity of methane is lower than He and Ne but higher than Ar and Kr, the conductivity of the mixture will be lower for He and Ne and higher for Ar and Kr by roughly ~1%. [71–73] The presence of decomposition products will have a similar effect depending on their molecular structure, with a particularly strong effect from $H_2$ due to its high conductivity.[72] Most importantly, plasma transport and the presence of graphitic particiles need to be included.[73] The description here is an oversimplification intended to address the differences in the thermal properties of the background inert gas.

Convective losses here are due to the gas removal by the flow:

$$q_{conv} = \frac{mnu_g C_p \Delta T}{L} = \frac{nu_g 5k_B \Delta T}{2L}, \tag{12}$$

where $k_B$ is the Boltzmann constant and $u_g$ is the flow speed of the gas in the tube (m/s). The given flow rate at room temperature and atmospheric pressure is $V_f$=50 scc/min, and the gas speed in the tube is.

$$u_g = \frac{V_f p_a T_g}{\pi R^2 T_a p} \tag{13}$$

where $p_a, p$ are the atmospheric pressure and gas pressure (2.5 torr), $T_a$ and $T_g$ are the atmospheric and gas temperatures, respectively, and R is the tube radius. For $T_g \sim 1000K$, $u_g \approx 10\ \frac{m}{s}$ and is higher for higher gas temperatures. With these estimates, $q_{conv} \ll q_{cond}$, the convection terms in the model are at least an order of magnitude lower than the conduction terms. We estimate that the flow rate required for more effective heat removal is of the order of liters per minute (not 50 sccm), which is typical for plasma jets.

Conduction appears to be the main heat-removal mechanism; hence, different gas conductivities may be responsible for the differences in the heating of the discharges in $CH_4$ with He, Ne, Ar, and Kr. He has the highest conductivity and hence cools the gas-suppressing thermal instabilities, whereas Ar and Kr cannot. Table 4 shows the estimated $\Delta T$ that can be supported by different gases compared with the experimental values of $\Delta T$ calculated from the measured wall temperature and Trot obtained by OES. This comparison confirmed the observed trends.

**Table 4.** Estimated $\Delta T$ from the axis to the outside surface of the tube and measured $\Delta T_g$ with $T_g$ from Table 2 and the outside surface temperature of 400 K .

|    | I, A | E, V/m | λ, W/m | max ΔTg, K | meas. ΔTg, K |            |
|----|------|--------|--------|------------|--------------|------------|
| He | 0.15 | 1500   | 0.1725 | 100        | 200          | He+5%$CH_4$ |
| Ne | 0.13 | 1500   | 0.0546 | 300        | 500          | Ne+5%$CH_4$ |
| Ar | 0.21 | 1500   | 0.033  | 1800       | 1100         | Ar+5%$CH_4$ |
| Kr | 0.16 | 1500   | 0.02   | 2300       | 1800         | Kr+5%$CH_4$ |

In addition to the simplifications specifically addressed above, such as the absence of molecular gases, $CH_4$ and its products, these estimations did not consider the presence of graphitic particles in the gas. Measurements in a similar RF reactor with a 1:1 mixture of Ar, $N_2$ and acetylene precursors used for carbon nanoparticle synthesis indicated that particle nucleation may also contribute significantly to gas heating.[74] This direction is left for future exploration.

**V.3 Products in the Diffuse and Contracted discharges**

In this section, we describe methane decomposition and product selectivity in the observed diffuse and contracted discharge modes.

Methane decomposition is achievable even without the addition of inert gas. Direct FTIR measurements showed up to 74% reduction in $CH_4$ concentration at 200 an input power. Although $H_2$ was not measured directly, OES confirmed strong H-atom emission and CH radical bands, indicating dissociation via the primary channel $CH_4$ → $CH_3$ + H. These results are consistent with earlier reports of RF discharges in pure methane, although requiring input powers >200 W for significant conversion.[1,75,76] The addition of inert gases substantially improves decomposition efficiency, reaching >90% at 150 W, consistent with previous studies on RF and dielectric barrier discharges.[76,77] In particular, with 5% $CH_4$ in Ar at 150 W, nearly complete decomposition (>99%) was achieved (Figure 8.b). This improvement correlates with the quenching of Ar metastable (Ar*), suggesting their involvement in dissociation and ionization pathways.[78,53,18,29,79] In addition, the necessary inclusion of metastable Ar* in the contraction model for direct dissociation and dissociative ionization of $CH_4$ also demonstrates the role of metastables in $CH_4$ decomposition. Additional information on the dynamics of Ar* metastables measured by LIF will be provided in a future study, self-consistent modeling of the discharge in diffuse and contracted modes is required to make detailed conclusions about the role of metastables.

Under our experimental conditions, the discharge existed in two distinct modes. In the diffuse mode, the plasma fills the tube uniformly, $T_{rot}$ <1000 K, and the plasma is non-thermal. OES shows strong CH and H emissions, while FTIR detects a wider range of hydrocarbon products ($C_2H_x$), characteristic of plasma-assisted, non-thermal decomposition.[42,53,75] Plasma proximity to the walls (at ~400 K) results in the extensive deposition of amorphous carbonaceous material on the tube walls.

In contrast, in contracted mode, observed in Ar/CH$_4$ and Kr/CH$_4$ mixtures (<5% CH$_4$), the discharge collapses into a hot axial filament a few mm in diameter, with $T_{rot}$ >1500 K in Ar and >2000 K in Kr (Table 2). Similar experiments (Nikhar et al.[12]) suggested that such filaments can sustain radial temperature gradients of >1000 K, confining the highest plasma density and energy deposition to the filament center. FTIR analysis confirmed nearly complete destruction of CH$_4$ (up to 99.7%) and product selectivity strongly favoring acetylene ($C_2H_2$), as expected under near-thermal equilibrium conditions at ~2000 K.[3,80–82] In this regime, no wall deposits were observed; instead, solid carbon was collected downstream on the filter, consistent with in-flow particle nucleation and growth.

The difference in the solid product morphology between the two modes was significant. Diffuse discharges yielded amorphous glassy deposits associated with rapid quenching of the cooled walls. In contrast, contracted discharges produce graphitized carbon nanomaterials, including onion-like structures, which is consistent with high-temperature annealing. The correlation between crystallinity and filament temperature mirrors earlier findings that increasing the treatment temperature (~2100 K) enhances graphitization, whereas excessive heating (>2500 K) may instead lead to evaporation.[10,11] Therefore, solid carbon formation depends on filament conditions, which are in turn dictated by the plasma regime and inert gas properties. It is also important to mention that particle heating can be provided by vibrationally excited molecules. Significant contributions of vibrationally excited N$_2$ to the nanoparticle temperature were demonstrated in [74].

The confinement of decomposition and nucleation to the contracted filament can be understood by comparing the molecular and particle transport timescales with gas residence time. At the experimental flow rates, the average gas residence time $t_g = \frac{L}{u_g} \approx 3 - 5\ ms$ with the gas speed estimated by (13). For the neutrals such as CH$_4$, C$_2$, and C$_2$H$_2$ in Ar, Chapman–Enskog estimates result in diffusion coefficients $D_g \approx 400 - 600\ cm^2 s^{-1}$, giving diffusion times to the wall, $t_d \sim \frac{R^2}{D_g} \approx 0.4\ ms$, much shorter than the residence time.[83,84] Hence, neutrals have enough time to diffuse to the walls, and neutral gas as expected, fills the entire tube. In contrast, carbon nanoparticles undergo many collisions with gas atoms or molecules due to their size $d \approx 20 - 50\ nm$ (Figure 10) and hence diffuse much slower. The diffusion coefficient is as defined in[85]: $D_p \approx \frac{(kT)^{1.5}}{6pd^2\sqrt{m_{Ar}}}$, which gives an estimated particle diffusion rate, $D_p \sim 0.01 - 0.2\ cm^2 s^{-1}$, corresponding to diffusion times on the order of seconds.[85,9] Therefore, neutral gas fills the tube, while nanoparticles formed in the gas phase close to the tube axis remain confined to the gas flow.

Because the convection flow is much faster than the particle diffusion time, the particles are carried downstream and collected on the filter.

A simple model in Sec. V.1 shows that the charged species recombine rapidly within the contracted filament volume, further limiting transport to the walls. These combined effects explain why wall deposits are absent in contracted discharges, and why solid products are collected almost exclusively in the downstream flow.

Together, these results show that the discharge mode controls both the gas- and solid-phase chemistry of the methane decomposition. Diffuse discharges, typical of He/CH$_4$, Ne/CH$_4$, or pure CH$_4$, promote non-thermal decomposition pathways with incomplete CH$_4$ conversion and amorphous wall deposits. Contracted discharges in Ar/CH$_4$ and Kr/CH$_4$, however, provide the high temperature and density conditions required for nearly complete CH$_4$ destruction and graphitized solid carbon production. The formation of contracted filaments depends on both the electronic and thermal properties of the added inert gas, including metastable excitation energies, quenching pathways, and thermal conductivity. These findings establish a clear link between plasma regime, gas composition, and carbon nanostructure formation, highlighting the potential for controlled synthesis under well-defined plasma conditions.

**V.5. A brief note on striations**

Our experiments show that the discharge mode depends on the background inert gas and that there is a correlation between the instabilities that are seen in the fast frame images for the pure gases and the contraction that occurs with the addition of CH$_4$. This brings attention to striation instabilities. [40]Here again gas heating can drive discharge instabilities by coupling the temperature, density, and electric field. At constant pressure, a temperature increase reduces the particle density, increases the reduced field (E/N), and thereby increases the electron temperature and density. The resulting current growth enhances ohmic heating (jE), further increasing the temperature and creating a positive feedback. [40]showed that an axial gas temperature rise of only 10–20% can trigger striations.[40] The strong dependence of ionization frequency on $T_e \propto E/N \propto T$ accelerates the instability. Efficient heat removal increased this threshold.

Experimentally, we confirm that in Ar gas, the discharge is uniform at a low input power (~15 W), whereas in He, the uniformity persists above ~150 W. Although these powers do not directly correspond to the heating or cooling estimates above, the trend is consistent with the thermal properties of He and Ar, with He being much more effective at maintaining non-thermal conditions in the discharge and suppressing heating instabilities. Recent studies on Ar discharges confirm the thermoelectric origin of striations for similar pR values, although at higher pressures, metastable pooling and stepwise ionization also contribute.[86,87,88,89]

We need to stress that we do not refer to ionization-thermal instability leading to contraction and only mention striations in pure inert gases here, because the choice of background gas appears to be very important for product selectivity under similar conditions.

## VI. Conclusions

The RF discharge in methane with the addition of an inert gas (2.5 torr) in a flow tube exhibits two modes: a diffuse, non-thermal mode with $T_{gas}$ <1000 K and a contracted mode, where a hot $T_{gas}$ >1500 K axial filament is unstable, moving around the tube.

The degree of decomposition of methane at the same input power increases with the addition of inert gases, He, Ne, Ar, and Kr. Decomposition is also effective in the diffuse non-thermal mode. The metastable Ar concentration decreased with increasing $CH_4$ admixture. These results provide evidence that non-thermal channels, electrons, and metastable states can facilitate effective decomposition. Laser-induced fluorescence measurements of metastable Ar states are currently underway and will be reported in the near future.

The decomposition products were different for the diffuse and contracted modes. In the diffuse mode, the decomposition appears to follow the non-thermal decomposition scheme, with strong OES signals from CH and H, lower degrees of decomposition, FTIR signals from a wider variety of gases, and copious amorphous deposits on the tube walls. In the contracted mode, the discharge contracts into a hot filament (>1500 K), where methane pyrolysis is favored with acetylene as the primary gas product and graphitized solid carbon collected on an in-flow filter. The tube walls remain clear. The degree of crystallization and its dependence on the plasma parameters will be investigated in the future.

The flow reactor in this study can be used as a flexible system that allows tunability of the desired gaseous and solid products.

**Acknowledgements:**

This work was supported by the U.S. Department of Energy through EERC (contract DE-AC02-09CH11466). We thank Yiguang Ju (EERC PI, PU) for fruitful collaboration, Stanislav Musikhin and Hengfei Gu (PU) for solid material analysis, SEM, TEM/EDS, and Aleksandr Merzhevsky for technical assistance.

**Supplementary Information:**

**OES C$_2$ Swan band fitted spectra using OES Toolbox:**

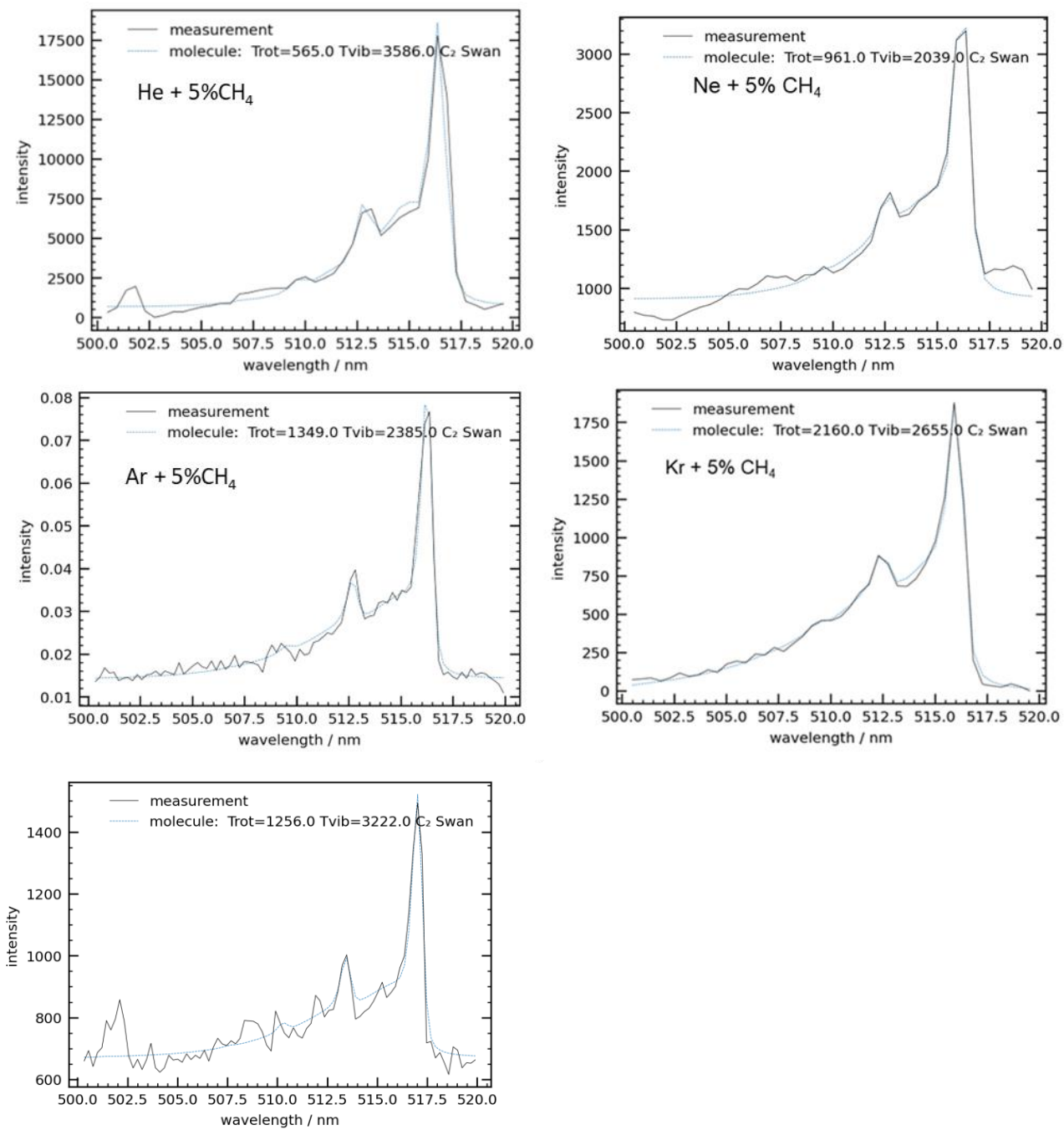

**Figure S1.** C$_2$ ro-vibrational band fitted using OES Toolbox. Rotational and vibrational excitation temperatures are the fitting parameters in the model. The sample results for the He, Ne, Ar, and Kr with 5% CH4 are given for the RF 13.56 MHz input power of 100 W, 2.6 torr, and 50 sccm total flow rate.

## Ar* metastables using OES Branching method:

**Table S1.** The upper and lower state chosen for the Branching method used to estimate the average concentration of Ar* metastable states.

| state upper | state lower | wavelength nm | Einstein Coefficient A, $s^{-1}$ |
|---|---|---|---|
| 2p1 | 1s4 | 667.73 | 2.36E+05 |
|  | 1s2 | 750.39 | 4.50E+07 |
| 2p2 | 1s5 | 696.54 | 6.40E+06 |
|  | 1s4 | 727.29 | 1.83E+06 |
|  | 1s3 | 772.42 | 1.17E+07 |
| 2p3 | 1s5 | 706.72 | 3.80E+06 |
|  | 1s4 | 738.4 | 8.50E+06 |
| 2p4 | 1s5 | 714.7 | 6.30E+05 |
|  | 1s4 | 747.12 | 2.20E+04 |
|  | 1s3 | 794.82 | 1.86E+07 |
| 2p6 | 1s5 | 763.51 | 2.45E+07 |
|  | 1s4 | 800.62 | 4.90E+06 |

**Hydrocarbon Concentrations in the Gas Ouput measured by FTIR absorption spectroscopy:**

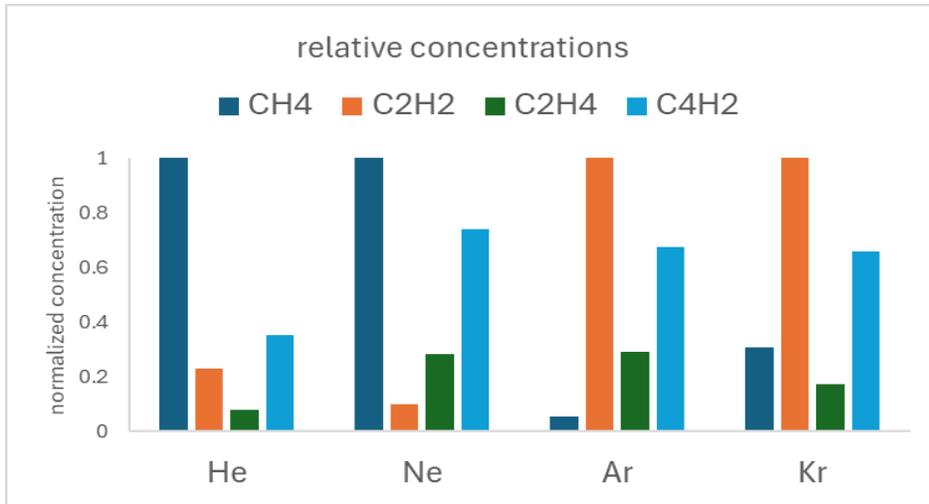

**Figure S2.** Hydrocarbon concentrations in the output gas normalized to the highest concentration.